\documentclass[accepted,specialissue]{melba}
\usepackage{mwe} % to get dummy images, only for the example
\usepackage{amsmath,amsfonts}
\usepackage{graphicx}
\usepackage{subfig}
\melbaid{2026:014}  % This is provided upon by the publishing editor
\doi{https://doi.org/10.59275/j.melba.2026-252c}
\melbaauthors{Chengze Ye, Linda-Sophie Schneider, Yipeng Sun, Mareike Thies, Siyuan Mei, Paula Andrea Pérez-Toro, Siming Bayer, and Andreas Maier}  % Note: this one is also used to set the pdf 'authors' metadata
\email{chengze.ye@fau.de}
\volume{2026}
\firstpageno{284}  % Communicated by the publishing editor
\melbayear{2026}  % The publication year
\datesubmitted{2025-07-15}  % Date submitted to MELBA: mm/yyyy
\datepublished{2026-07-02}  % Today's date: mm/yyyy

\melbaspecialissue{MELBA–BVM 2025 Special Issue}
\melbaspecialissueeditors{Andreas Maier, Thomas Deserno, Heinz Handels, Klaus Maier-Hein, Christoph Palm, Thomas Tolxdorff, Katharina Breininger}

\ShortHeadings{Robustness and Stability Analysis of Differentiable SV-FBP}{Ye et al.}

\title{Robustness and Stability Analysis of Differentiable Shift-Variant FBP for Cone-Beam CT under Challenging Acquisition Settings}

\author{
	\name Chengze Ye \orcid{0009-0001-0221-506X},
	\name Linda-Sophie Schneider \orcid{0009-0002-6195-9859},
    \name Yipeng Sun \orcid{0009-0009-8587-8894},
    \name Mareike Thies,\orcid{0000-0002-1364-4337}
    \name Siyuan Mei \orcid{0009-0002-1136-2501},
    \name Paula Andrea Pérez-Toro \orcid{0000-0002-2727-2116},
    \name Siming Bayer \orcid{0000-0003-2874-4805}, 
    \name Andreas Maier \orcid{0000-0002-9550-5284}
}
\affiliations{% <- trailing '%' to avoid unwanted indent
	\addr Pattern Recognition Lab, Friedrich-Alexander-Universität Erlangen-Nürnberg, Erlangen, Germany \\
}

\abstract{The differentiable shift-variant filtered backprojection (SV-FBP) framework enables data-driven estimation of redundancy weights for cone-beam CT reconstruction under general source trajectories, removing the need for analytically derived weighting schemes. In this work, we present a systematic study of the robustness and adaptability of differentiable SV-FBP under challenging acquisition settings. We show that the framework remains stable across highly irregular and discontinuous trajectories, indicating that reconstruction performance is largely insensitive to trajectory ordering or continuity. Instead, the spatial distribution of sampling points plays a more dominant role. Under sparse-view conditions, differentiable SV-FBP achieves competitive reconstruction quality while providing an order-of-magnitude reduction in computation time compared to iterative reconstruction methods at moderate sampling densities. However, we identify a clear transition regime under severe undersampling, where the absence of iterative data consistency leads to performance degradation. Furthermore, we demonstrate that the framework remains applicable to non-planar multi-isocenter geometries, such as Lissajous-saddle trajectories, without requiring architectural modifications. These findings provide new insights into the behavior and limitations of the differentiable SV-FBP model and highlight it as a flexible and efficient solution for non-standard and robotic CBCT acquisition scenarios.}


\begin{document}

% top matter
\twocolumn[\maketitle]
% comment the preceedings and uncomment the following if the authors list + abstract is longer than one page
% \maketitle
% \twocolumn

% Introduction (or first section)
% \rule{\textwidth}{1pt}
\section{Introduction}

\enluminure{C}{one} \emph{beam} Computed tomography (CBCT) is a widely used imaging modality in operating rooms, enabling the acquisition of three-dimensional (3D) images within a single rotation at a relatively low radiation dose. Its compact design and rapid acquisition capability make it particularly well suited for image-guided interventions such as angiography and spine surgery. Nevertheless, maintaining high image quality in these dynamic settings remains challenging, particularly in the presence of metal implants, which frequently cause artifacts and reconstruction errors.

In recent years, the advent of robotic C-arm systems has enabled the exploration of scanning trajectories beyond conventional circular scans. These systems can follow non-circular paths tailored to patient anatomy and specific clinical requirements~\citep{hatamikia2022source}. For example, sinusoidal trajectories and other carefully designed acquisition paths have been shown to significantly reduce metal artifacts~\citep{gang2020metal, gang2020non}. Optimized trajectories can also improve image quality while reducing the number of required projections~\citep{herl2020scanning}. Saddle trajectories, in particular, have proven effective in mitigating cone-beam artifacts and enabling reliable reconstruction even in the presence of axial truncation~\citep{pack2004investigation}.

However, non-circular trajectories also pose substantial challenges for image reconstruction. The Feldkamp-Davis-Kress (FDK) algorithm~\citep{feldkamp1984practical}, one of the most widely used reconstruction methods, assumes a circular orbit and therefore degrades markedly when this assumption is violated. Iterative approaches such as model-based iterative reconstruction (MBIR)~\citep{liu2014model} can achieve high reconstruction accuracy for more general trajectories, but they typically require substantial computational resources. In addition, the design of suitable regularization terms for MBIR is often challenging, which further limits its practical applicability.

To address this issue, Defrise and Clack proposed a shift-variant filtered backprojection (FBP) algorithm that adapts to varying scanning geometries~\citep{defrise1994cone}. Although this approach considerably reduces reconstruction time compared with iterative methods, it still requires the estimation of trajectory-dependent weights. Importantly, the analytical form of these weights depends on the derivative of the scanning trajectory, which makes the method difficult to apply to discontinuous or piecewise-defined acquisition paths.

In our previous work~\citep{ye2024deep}, we introduced a differentiable reconstruction framework based on shift-variant filtered backprojection and known-operator learning, in which redundancy weights are estimated in a data-driven manner. The initial conference version established the core formulation, while subsequent studies extended the evaluation to more challenging trajectory settings, including continuous non-circular scans~\citep{ye2025draco} and discontinuous random trajectories~\citep{ye2025learned}.

This paper is intended as a systematic extension of these earlier studies rather than as a new reconstruction architecture. While prior work demonstrated the feasibility of differentiable SV-FBP under selected trajectory classes, several practical questions remain open, including the sensitivity of the framework to trajectory discontinuity and sampling order, the sparse-view conditions under which its performance begins to deteriorate relative to iterative reconstruction, and whether the method remains applicable beyond fixed-isocenter acquisition geometries. To address these questions, we retain the core differentiable SV-FBP architecture and extend the evaluation in three directions:

\begin{itemize}
\item \textbf{Robustness under discontinuous trajectories.} In addition to random and random nearest-neighbor reordered trajectories, we include random farthest-neighbor reordered trajectories to assess stability under more aggressive discontinuities and spatial irregularity. Our results show that reconstruction performance is largely insensitive to trajectory ordering or continuity and is instead governed more strongly by the spatial distribution of sampling points.

\item \textbf{Sparse-view performance and limitation analysis.} We evaluate the framework under varying levels of projection sparsity and identify a transition regime in sparse-view reconstruction, where differentiable SV-FBP remains competitive at moderate sampling densities but degrades under severe undersampling because it lacks iterative data consistency.

\item \textbf{Evaluation beyond fixed-isocenter scans.} To address the restriction of earlier robustness studies to fixed-isocenter trajectories, we further evaluate the framework on a Lissajous-saddle trajectory with a continuously varying rotation center, thereby examining its applicability in a multi-isocenter acquisition setting.
\end{itemize}

The remainder of this paper is organized as follows. Section~\ref{sec:2} reviews related work. Section~\ref{sec:3} introduces the theoretical background and describes the source trajectory designs used in the experiments. Sections~\ref{sec:4} and~\ref{sec:5} present the experimental setup and results, including trajectory robustness and sparse-view reconstruction analyses. Section~\ref{sec:6} discusses the main findings, and Section~\ref{sec:7} concludes the paper and outlines possible directions for future work.

%%%%%%%%%%%%%%%%%%%%%%%%%%%%%%%%%%%%%%%%%%%%%%%%%%%%%%%%%%%%%%%%%%%%%%%%%%%
% Related works
%%%%%%%%%%%%%%%%%%%%%%%%%%%%%%%%%%%%%%%%%%%%%%%%%%%%%%%%%%%%%%%%%%%%%%%%%%%
% Make sure to put your work into context and include apporpriate citations.
% We do not have limits on citation counts.
\section{Related Work}
\label{sec:2}

In this section, we briefly review representative reconstruction methods most relevant to the present study, including conventional CBCT reconstruction methods and deep learning-based approaches for non-circular trajectories.

\subsection{Conventional CBCT Reconstruction Methods}

% CBCT的早期重建方法多以解析型算法为主，代表性方法为Feldkamp–Davis–Kress（FDK）算法。FDK方法具备高计算效率，在理想的圆轨迹与均匀采样条件下可实现快速的三维重建。这种滤波反投影类型的算法，滤波操作和数据采集可以同时进行，大大提高在实际应用效率。然而，该算法无法适应轨迹变化等复杂场景，在非圆轨迹条件下易产生明显伪影，限制了其在robotic C臂系统中的应用。

Early CBCT reconstruction methods primarily relied on analytic algorithms, with the Feldkamp–Davis–Kress (FDK) algorithm~\citep{feldkamp1984practical} being the most widely used. FDK has been demonstrated to offer high computational efficiency and enable fast 3D reconstruction under ideal circular trajectories with uniform angular sampling. As a FBP-type method, it is computationally efficient and well suited for fast 3D reconstruction in practice. However, its performance is significantly degraded by non-circular trajectories, which limits its applicability in robotic C-arm systems.

% 为重建非圆形轨迹，模型基迭代重建（Model-Based Iterative Reconstruction, MBIR）方法被提出。MBIR算法需要多次的forward and backward projection computations来。尽管其重建精度高，但因其高度迭代性，计算开销大、重建速度慢，不适用于临床实时成像。 此外用于重建的正则项通常很难去设计,也大大加剧了.。。

To handle non-circular trajectories, model-based iterative reconstruction (MBIR) methods~\citep{liu2014model} have been widely employed. These methods involve multiple forward and backward projection computations with the objective of iteratively reconstructing the image. Although they offer high reconstruction accuracy, their heavy computational burden and slow convergence limit their suitability for real-time clinical applications. Furthermore, designing appropriate regularization terms is often non-trivial and further increases implementation complexity.

% In terms of theoretical modeling, Grangeat's method established the relationship between cone-beam projection data and a function associated with the 3D Radon transform of the object, providing a foundational basis for reconstruction from non-circular trajectories~\citep{grangeat1991mathematical}. Based on this relationship, a methodology for the exact reconstruction of arbitrary orbits was proposed. However, the reconstruction process is rendered relatively slow and memory-intensive by the necessity of constructing a high-dimensional intermediate function matrix.
Grangeat's method established the relationship between cone-beam projection data and a function associated with the 3D Radon transform of the object, providing an important theoretical basis for reconstruction from non-circular trajectories~\citep{grangeat1991mathematical}. Based on this formulation, exact reconstruction for general source trajectories becomes possible in principle. In practice, however, the need to construct a high-dimensional intermediate function makes the reconstruction process relatively slow and memory-intensive.

% 基于result of Grangeat，Defrise 和 Clack 提出Shift-Variant Filtered Backprojection（SV-FBP）算法，引入与轨迹相关的冗余权重，用以解决several cone beam projection supply the same values in Radon domain的问题。该算法不再需要3D rebinning step，可以实现非圆形估计快速重建
Building on Grangeat's formulation, Defrise and Clack proposed the Shift-Variant Filtered Backprojection (SV-FBP) algorithm, which introduces trajectory-dependent redundancy weights to address the issue that multiple cone-beam projections may contribute redundant information in the Radon domain. The elimination of the 3D rebinning step enables rapid reconstruction under non-circular trajectories~\citep{defrise1994cone}.

\subsection{Deep Learning-Based Reconstruction Methods}

In recent years, deep learning has been increasingly applied to CBCT reconstruction tasks~\citep{koetzier2023deep}. End-to-end networks such as AUTOMAP~\citep{zhu2018image} and iRadonMAP~\citep{he2020radon} learn a direct mapping from projection data, typically acquired along the circular trajectory, to reconstructed images. These methods have shown promising results in low-dose CT applications and artifact reduction. However, when applied to projection data from non-circular trajectories, the variability in the input increases significantly, requiring larger network capacity to effectively model the complex mapping.

Additionally, recent AI-based approaches such as Neural Radiance Fields (NeRF)~\citep{wang2024neural}, implicit neural representations (INR)~\citep{molaei2023implicit}, and 3D Gaussian Splatting~\citep{li2023sparse,lin2024learning,wu2024differentiable} have achieved remarkable progress in 3D scene reconstruction from natural images. Their strong representational power makes them theoretically appealing for CBCT reconstruction under non-circular trajectories. However, these methods typically require scene-specific optimization or retraining for each new object, often taking several minutes to hours, which is impractical for time-sensitive clinical applications. Moreover, these methods are generally considered “black-box” models with limited interpretability, raising concerns regarding their transparency and reliability in medical imaging. 

To improve interpretability and incorporate physical priors, the Known Operator Learning framework has been proposed, which involves integrating known operators as prior knowledge into machine learning models~\citep{maier2019learning}. This approach reduces model complexity by minimising the number of trainable parameters, thereby reducing the amount of training data required and the maximal error bounds. This precision learning strategy has already been successfully applied in medical imaging and has shown promising results~\citep{wurfl2018deep, syben1807deriving, sun2025learning}.

% A methodological, model, or similar section often comes here.
\section{Methods}
\label{sec:3}
\subsection{Shift-Variant FBP Algorithm}
The shift-variant FBP algorithm, proposed by Defrise and Clack~\citep{defrise1994cone}, has been developed to reconstruct CBCT data with a specific trajectory. The reconstruction formula is as follows:
\begin{align*}
f(x)&=\int_{ \Lambda}^{} d\lambda\int_{S^2/2}^{}d\theta-\frac{1}{4\pi^2}\mid a'(\lambda)\cdot\theta\mid\\
&\frac{1}{n(\theta, \lambda)}
\times\delta' ((x-a(\lambda))\cdot \theta)S(\theta, \lambda ).\tag{1}
\label{3684-eq:myequation1}
\end{align*}

In this formula, $\lambda \in \Lambda$ refers to the parameter defining the source position $a(\lambda)$, while $\theta \in S^2$, where $S^2$ represents the set of all unit vectors in $\mathbb{R}^3$. The function $S(\theta, \lambda )$ is Grangeat's intermediate function, which connects cone-beam projections to the first derivative of the Radon transform. Additionally, $n(\theta, \lambda)$ denotes the number of intersections between the trajectory $a(\Lambda)$ and the plane orthogonal to $\theta$, passing through the point $a(\lambda)$ along the trajectory.

\begin{figure*}[] % [b]
\centering
\includegraphics[width=7.5in]{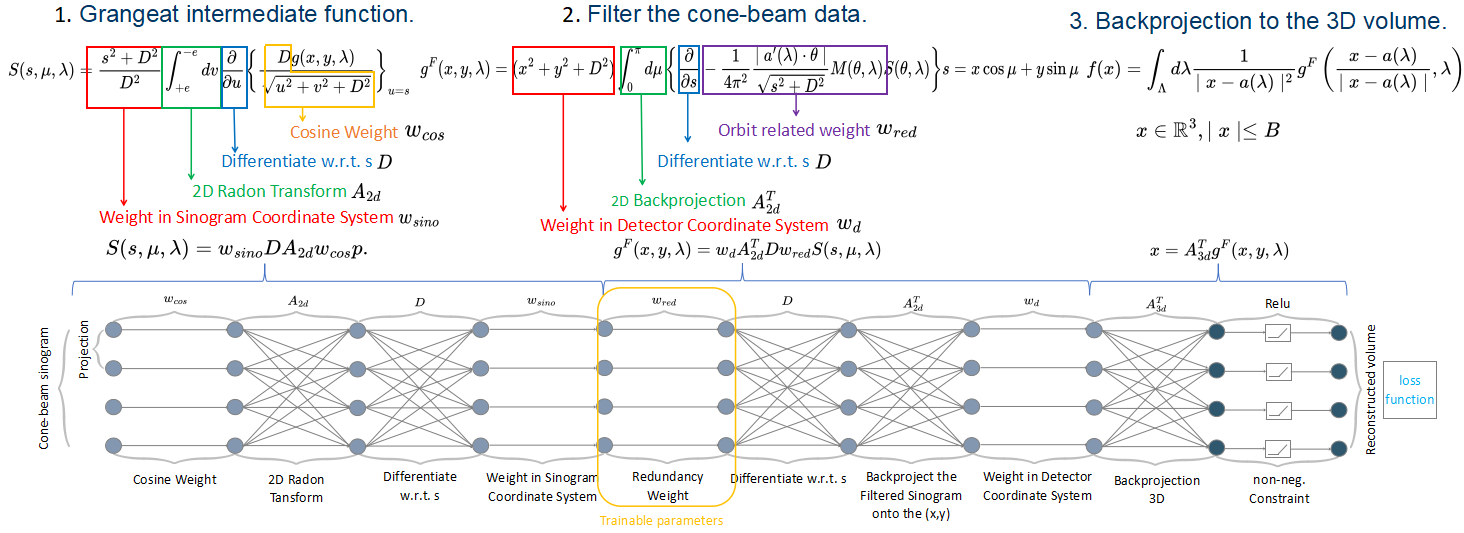}
\caption{Differentiable shift-variant FBP model architecture.}
\label{3684-network}
\end{figure*}

In equation~\eqref{3684-eq:myequation1}, the function $n(\theta, \lambda)$ exhibits discontinuities, which can result in artifacts during discrete implementations. In order to resolve this issue, Defrise and Clack replaced the crofton symbol $\frac{1}{n(\theta, \lambda)}$ with a smooth, differentiable function:
\begin{align*}
M(\theta,\lambda) = \frac{\mid a'(\lambda)\cdot\theta\mid^m c(\lambda)}{\sum_{\alpha=1}^{n(\theta, \lambda)} \mid a'(\lambda_\alpha)\cdot\theta\mid^m c(\lambda_\alpha)
}, \tag{2}
\label{3684-eq:myequation2}
\end{align*}
where $m$ is a positive integer greater than 2, $c(\lambda)$ is a smooth function that is equal to one across almost the entire interval $\Lambda$, except near the interval boundaries, where it tends toward 0. 

Nevertheless, the practical calculation of equation~\eqref{3684-eq:myequation2} presents a number of challenges. For instance, selecting an appropriate parameter $m$ for a given trajectory geometry requires experiments. Furthermore, computing the gradient of the trajectory $a'(\lambda)$ for discontinuous orbits is impossible. This limitation is one of the main motivations for the data-driven formulation introduced in the next subsection.

To enable practical implementation, Defrise and Clack explicitly rewrote Equation~\eqref{3684-eq:myequation1} for the case in which cone-beam data are acquired on a flat planar detector. The geometry is described in the detector coordinate system, where each projection is recorded using detector coordinates $(u,v)$, and $D$ denotes the distance from the X-ray source to the detector. The reconstruction process can be decomposed into three steps. First, the cone-beam projections are transformed into Grangeat’s intermediate function via:
\begin{align*}
S(s,\mu,\lambda)=\frac{s^2+D^2}{D^2}\int_{+e}^{-e}dv\frac{\partial }{\partial u}\left\{\frac{Dg(x,y,\lambda)}{\sqrt{u^2+v^2+D^2}}\right\} _{u=s}.\tag{3}
\label{eq:myequation3}
\end{align*}

Then, within the sinogram domain parameterized by $(s,\mu)$, Grangeat’s intermediate function is weighted and differentiated before being mapped back to the detector coordinate system:

\begin{align*}
g^F(x,y,\lambda)&=(x^2+y^2+D^2)\int_{0}^{\pi} d\mu\left\{\frac{\partial }{\partial s}\frac{S_1(s,\mu,\lambda)}{\sqrt{s^2+D^2}}\right\} \\
s&=x\cos\mu+y\sin\mu \\ 
S_1(s,\mu,\lambda)&=-\frac{1}{4\pi^2}\mid a'(\lambda)\cdot\theta\mid M(\theta,\lambda)S(\theta,\lambda).\tag{4}
\label{eq:myequation4}
\end{align*}

Finally, the filtered data are backprojected into the image volume through a weighted backprojection step~\eqref{eq:myequation5}, yielding the final reconstruction result without requiring 3D rebinning.

\begin{align*}
f(x)&=\int_{\Lambda}^{}d\lambda \frac{1}{\mid x- a(\lambda)\mid ^2} g^F(\frac{x- a(\lambda)}{\mid x- a(\lambda)\mid}, \lambda)\tag{5}
\label{eq:myequation5}
\end{align*}

\begin{figure*}[]
    \centering
    \subfloat[]{\includegraphics[width=2.3in]{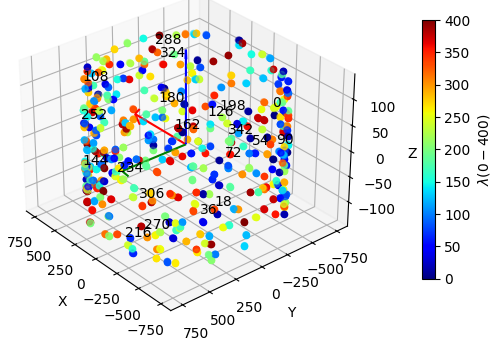}}
    \subfloat[]{\includegraphics[width=2.3in]{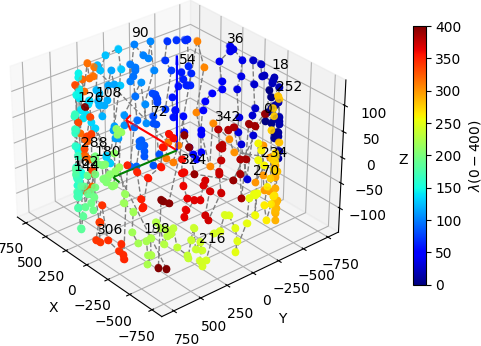}}
    \subfloat[]{\includegraphics[width=2.3in]{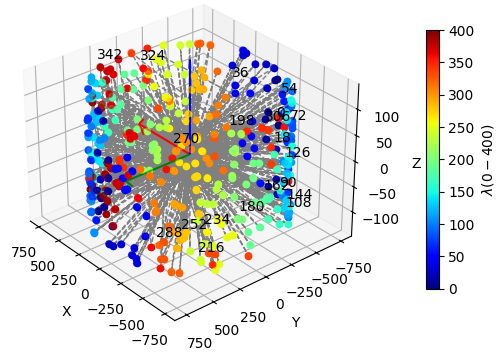}}
    \caption{Illustration of the three source trajectory types: (a) Random Trajectory (RT), (b) Random Nearest-Neighbor Reordered (RNNR), and (c) Random Farthest-Neighbor Reordered (RFNR).}
    \label{fig:trajectories}
\end{figure*}

\subsection{Differentiable Shift-Variant FBP Neural Network}

Building on the analytical SV-FBP formulation, our previous work~\citep{ye2024deep} introduced a differentiable reconstruction model for CBCT under general source trajectories. Following the principle of known operator learning, the original reconstruction pipeline is reformulated as a neural network in which the analytical operators are retained explicitly and only the trajectory-dependent weighting term is learned from data. This leads to the following differentiable representation:
\begin{align*}
f(x)=A_{3d}^Tw_{d}A_{2d}^TDw_{red}w_{sino}DA_{2d}w_{cos}g(x,y,\lambda)\tag{6}
\label{3684-eq:myequation6}
\end{align*}
as illustrated in Figure~\ref{3684-network}.

In this formulation, a cosine weight $w_{cos}$ is first applied to the projection data $g(x,y,\lambda)$, followed by a 2D Radon transform $A_{2d}$. A differentiation operator $D$ and the weighting term $w_{sino}$ are then used in the sinogram domain to compute Grangeat's intermediate representation. The trajectory-dependent factors in Eq.~\eqref{3684-eq:myequation1}, namely $\frac{1}{n(\theta,\lambda)}$ and $\mid a'(\lambda)\!\cdot\!\theta \mid$, are absorbed into a redundancy weight $w_{red}$, which is modeled as a trainable parameter. After applying a second differentiation, the data are mapped back to the detector domain by $A_{2d}^T$, weighted by $w_d$, and finally reconstructed into the image volume through the 3D backprojection operator $A_{3d}^T$.

A key property of this formulation is that only the redundancy weights are learned, while all remaining components follow a differentiable implementation of the original SV-FBP algorithm~\citep{defrise1994cone}. This substantially reduces the number of trainable parameters and preserves the interpretability of the reconstruction pipeline, while enabling data-driven adaptation to different trajectory geometries.

\subsection{Trajectory Design}
\label{sec:trajectory_types}

To assess the robustness of our model under different acquisition conditions, we consider two complementary classes of trajectories. The first class comprises fixed-isocenter randomized trajectories, which are designed to challenge the reconstruction with respect to sampling continuity and angular ordering. The second class consists of a multi-isocenter trajectory, which extends the evaluation to geometries in which the rotation center varies continuously throughout the scan. Together, these two classes allow us to separately examine the effects of sampling irregularity and geometric non-stationarity.

\subsubsection{Fixed-Isocenter Trajectories}
\label{sec:fixed-iso}

The fixed-isocenter trajectories are constructed under the assumption that the object lies at the center of a spherical acquisition surface and that all source positions are located on a shell of fixed radius $R$. The source position is expressed in spherical coordinates as
\begin{align*}
S_i(\theta_i, \phi_i) = \left(\begin{array}{c}x\\ y\\ z\end{array}\right)=  \left(\begin{array}{c}R\cos\theta_i \cos\phi_i\\ R\sin\theta_i \cos\phi_i\\ R\sin\phi_i\end{array}\right) ,\tag{7}
\label{3684-eq:myequation7}
\end{align*}
where $\theta_i\in [0,2\pi)$ denotes the in-plane rotation angle (gantry angle), and $\phi_i \in [-\phi_{\max}, \phi_{\max}]$ denotes the out-of-plane tilt. In our experiments, $R=750$\,mm and $\phi_{\max} = 10^\circ$ were chosen to reflect the mechanical constraints of the Artis zeego system. For each source position, the detector is oriented such that its central axis points toward the isocenter. The horizontal and vertical detector axes are defined to form a right-handed orthonormal coordinate system, ensuring a consistent projection geometry across all views.

We construct three variants of this trajectory class, illustrated in Fig.~\ref{fig:trajectories}:
\begin{enumerate}
    \item \textbf{Random Trajectory (RT)}: $\theta_i$ and $\phi_i$ are sampled independently from uniform distributions:
    \begin{equation}
    \theta_i \sim \mathcal{U}(0, 2\pi), \quad \phi_i \sim \mathcal{U}(-\phi_{\max}, \phi_{\max}). \tag{8}
    \end{equation}
    This produces a fully randomized source trajectory.

    \item \textbf{Random Nearest-Neighbor Reordered (RNNR)}: Given a random set $\{\mathbf{S}_i\}_{i=1}^N$, the source positions are reordered to minimize jumps between consecutive views:
    \begin{align}
        i_{k+1} &= \underset{j \notin \{i_1, \dots, i_k\}}{\arg\min} \|\mathbf{s}_j - \mathbf{s}_{i_k}\|. \tag{9}
    \end{align}
    Starting from an initial source position, this produces a relatively continuous traversal while preserving the underlying random sampling locations.

    \item \textbf{Random Farthest-Neighbor Reordered (RFNR)}: In contrast, the source positions are reordered to maximize jumps between consecutive views:
    \begin{equation}
    i_{k+1} = \underset{j \notin \{i_1, \dots, i_k\}}{\arg\max} \|\mathbf{s}_j - \mathbf{s}_{i_k}\|. \tag{10}
    \end{equation}
    This yields the most aggressive discontinuity pattern among the fixed-isocenter trajectories and therefore represents the most challenging case in this trajectory class.
\end{enumerate}

\subsubsection{Multi-Isocenter Trajectory}
\label{sec:lissajous-def}

While the fixed-isocenter trajectories above probe the effect of sampling irregularity, they do not alter the underlying acquisition geometry. To extend the evaluation beyond fixed-isocenter scans, we introduce a Lissajous-saddle trajectory in which the effective rotation center varies continuously throughout the acquisition. This produces a non-planar and asymmetric source path that cannot be approximated by a conventional circular scan.

Specifically, the isocenter follows a smooth 3D Lissajous curve:
\begin{align*}
C(\theta) = \bigl(A_x\sin(a\theta),\; A_y\sin(b\theta+\delta),\; A_z\cos(c\theta)\bigr)^\top,
\end{align*}
where $A_x, A_y, A_z$ are the displacement amplitudes along the three axes, $a, b, c$ are mutually coprime integers controlling the Lissajous frequencies, and $\delta$ is a phase offset that breaks planar symmetry. Simultaneously, the X-ray source executes a saddle-type out-of-plane motion around the instantaneous isocenter,
\begin{align*}
\phi(\theta) = \phi_{\max}\cos(f\theta),
\end{align*}
which gives the source position at projection angle $\theta$ as
\begin{align*}
\mathbf{S}(\theta) = C(\theta) + R\bigl(\cos\theta\cos\phi,\;\sin\theta\cos\phi,\;\sin\phi\bigr)^\top,
\end{align*}
where $R$ denotes the source-to-isocenter distance.

For each projection, the detector coordinate frame is constructed from the central-ray vector
\[
\mathbf{k} = \frac{C(\theta) - \mathbf{S}(\theta)}{\|C(\theta) - \mathbf{S}(\theta)\|},
\]
the horizontal vector
\[
\mathbf{h} = \frac{\mathbf{k} \times \mathbf{e}_z}{\|\mathbf{k} \times \mathbf{e}_z\|},
\]
with $\mathbf{e}_z$ replaced by $\mathbf{e}_x$ when $|\mathbf{k} \cdot \mathbf{e}_z| > 0.99$ to avoid degeneracy, and the vertical vector
\[
\mathbf{v} = \mathbf{k} \times \mathbf{h}.
\]
This defines a consistent detector reference frame despite the continuously shifting isocenter.

% The following section will present the experiments conducted to analyse the potential and robustness of this reconstruction pipeline across both trajectory classes.

\begin{table*}[]
\centering
\caption{Comparison of Image Quality Metrics for Different Trajectory Categories.}
%\resizebox{\textwidth}{!}{
\begin{tabular}{@{}lcccccc@{}}
\hline
\toprule

 & \multicolumn{4}{c}{\emph{Differentiable SV-FBP}}  & \emph{SV-FBP}\\
\midrule
\emph{Trajectory Type}& \emph{Seed} & \emph{MSE} $\downarrow$ & \emph{PSNR (dB)} $\uparrow$ & \emph{SSIM} $\uparrow$ \\ 
\cmidrule{1-5}

\multirow{6}{*}{\emph{RT}} & Seed0 & 0.1050 ± 0.0157 & 35.88 ± 1.27 & 0.9209 ± 0.0085 & \multirow{6}{*}{$N/A$} \\
 & Seed1 & 0.1029 ± 0.0170 & 36.08 ± 1.28 & 0.9240 ± 0.0056 \\
 & Seed2 & 0.1057 ± 0.0176 & 35.84 ± 1.24 & 0.9210 ± 0.0068 \\
 & Seed3 & 0.1020 ± 0.0167 & 36.15 ± 1.25 & 0.9264 ± 0.0059 \\
 & Seed4 & 0.1037 ± 0.0172 & 36.00 ± 1.24 & 0.9223 ± 0.0064 \\
 & \textbf{Overall} & \textbf{0.1039 ± 0.0169} & \textbf{35.98 ± 1.26} & \textbf{0.9229 ± 0.0070}\\
\hline
\multirow{6}{*}{\emph{RNNR}} & Seed0 & 0.1042 ± 0.0159 & 35.95 ± 1.26 & 0.9225 ± 0.0079& \multirow{6}{*}{$N/A$}\\
 & Seed1 & 0.1015 ± 0.0169 & 36.19 ± 1.28 & 0.9237 ± 0.0059\\
 & Seed2 & 0.1059 ± 0.0173 & 35.82 ± 1.25 & 0.9199 ± 0.0070\\
 & Seed3 & 0.1044 ± 0.0163 & 35.94 ± 1.24 & 0.9242 ± 0.0058\\
 & Seed4 & 0.1021 ± 0.0167 & 36.13 ± 1.24 & 0.9248 ± 0.0064\\
 & \textbf{Overall} & \textbf{0.1036 ± 0.0167} & \textbf{36.01 ± 1.26} & \textbf{0.9230 ± 0.0069}\\
\hline
\multirow{6}{*}{\emph{RFNR}}  & Seed0 & 0.1041 ± 0.0160 & 35.95 ± 1.25 & 0.9208 ± 0.0079 & \multirow{6}{*}{$N/A$}\\
 & Seed1 & 0.1023 ± 0.0172 & 36.13 ± 1.29 & 0.9248 ± 0.0056 \\
 & Seed2 & 0.1055 ± 0.0172 & 35.85 ± 1.26 & 0.9235 ± 0.0068 \\
 & Seed3 & 0.1023 ± 0.0165 & 36.12 ± 1.27 & 0.9259 ± 0.0063 \\
 & Seed4 & 0.1057 ± 0.0170 & 35.84 ± 1.24 & 0.9180 ± 0.0067 \\
 & \textbf{Overall} & \textbf{0.1040 ± 0.0168} & \textbf{35.98 ± 1.26} & \textbf{0.9226 ± 0.0067}\\
\bottomrule

\end{tabular}%}%
\label{tab:Trajectory}
\end{table*}

\section{Experimental Setup}
\label{sec:4}
\subsection{Simulation Geometry Configuration}
To ensure a consistent and realistic simulation environment, we replicated the geometry of the clinical Artis zeego C-arm CT system. The simulated detector operated in a $4 \times 4$ binning mode, yielding an effective resolution of $620 \times 480$ pixels with a detector spacing of 0.616~mm. The source-to-detector distance was set to 1200~mm, and the source-to-isocenter distance was set to 750~mm, consistent with the clinical system configuration. The gantry angle $\theta$ was sampled uniformly over $[0, 2\pi)$, and the maximum tilt angle was fixed at $10^\circ$.

For each scan, a 3D volume of size $128 \times 512 \times 512$ voxels with isotropic voxel spacing of 0.25~mm was reconstructed. Unless otherwise stated, 400 cone-beam projections were generated per scan, corresponding to the reference acquisition setting used throughout the experiments.

\subsection{Data Preparation}
\label{sec:data-preparation}

To evaluate the robustness of differentiable SV-FBP in a controlled setting, we followed the data generation framework established in our prior work~\citep{ye2025draco} and combined synthetic training data with a small real-patient inference set.

\paragraph{Synthetic data.}
Ground-truth volumes were generated by placing a random number of geometric primitives with varying shapes, positions, orientations, and densities into a voxel volume. In addition, cylinders with randomly sampled diameters were placed along the longitudinal axis to approximate the structural characteristics of the thoracic and abdominal regions. To reduce the influence of unrealistically sharp boundaries during training, a Gaussian filter was applied to each volume before forward projection.

\paragraph{Trajectory instances.}
For the fixed-isocenter trajectory classes (RT, RNNR, and RFNR), the gantry angle $\theta$ was sampled uniformly over $[0, 2\pi)$ and the tilt angle $\phi$ was sampled independently from $\mathcal{U}(-\phi_{\max}, \phi_{\max})$ for each projection, with $\phi_{\max}=10^\circ$. For each fixed-isocenter trajectory type, five independent random seeds were used to generate five distinct trajectory instances, resulting in 15 fixed-isocenter trajectories in total. This design enables a controlled analysis of trajectory-dependent variability and the effect of source ordering.

For the Lissajous-saddle trajectory, projection angles were distributed uniformly over the full rotation range, while the rotation center varied continuously according to the trajectory definition in Section~\ref{sec:lissajous-def}. The parameters used for this trajectory were $A_x=20$, $A_y=15$, $A_z=60$, $a=2$, $b=3$, $c=5$, $\delta=\pi/4$, and $f=2$.  

\paragraph{Training and validation data.}
For each trajectory instance, 30 cone-beam projection datasets were synthesized using PyroNN~\citep{syben2019pyro}. Among these, 24 were used for training and 6 for validation. The final sinograms were obtained by forward-projecting the synthetic ground-truth volumes using the corresponding cone-beam geometry of each trajectory instance.

\paragraph{Real-patient data.}
To complement the synthetic experiments with more realistic imaging content, we additionally used 5 patient scans from the publicly available Pancreatic-CT-CBCT-SEG dataset~\citep{hong2021breath}. These scans were forward-projected using the same acquisition geometry and used for inference-only evaluation.

\paragraph{Sparse-view settings.}
Sparse-view experiments were derived from the same full-view acquisition geometry by reducing the number of projections from 400 to 300, 200, and 100 views. This allows the effect of undersampling to be evaluated while keeping the underlying scan geometry fixed.

\subsection{Implementation Details}
\label{sec:implementation}

The neural network was implemented using the PyTorch 2.1.1 deep learning framework and trained on an NVIDIA A40 GPU. The network included specialized differentiable operators from the PyroNN library~\citep{syben2019pyro} for 2D Radon transforms, 2D backprojection, and 3D cone-beam backprojection, ensuring physical fidelity in the simulation and reconstruction pipeline.

A Gaussian filter was applied after the redundancy weight layer. This design helps smooth the weights, preventing spike-like responses in the filtered projections that would otherwise lead to reconstruction artifacts. The filter used a fixed kernel size of 121 and a standard deviation (sigma) of 20, chosen empirically to ensure sufficient smoothness without excessively blurring the weight function. The initialization of the redundancy weight layer was conducted using values drawn from a uniform distribution within the interval $[-1, 0]$, as we observed that the learned weights tend to exhibit a globally negative bias. This initialization provides a more suitable starting point and empirically accelerates convergence.

\subsection{Loss Function and Optimization}
\label{sec:loss}

The loss function used to guide training consists of two complementary terms: the mean squared error (MSE) loss $\mathcal{L}_{mse}$ and the structural similarity index measure (SSIM) loss $\mathcal{L}_{ssim}$. These were linearly combined by a fixed weighting factor $\gamma$ of $5 \times 10^{-3}$, promoting both pixel-wise accuracy and perceptual fidelity in the reconstructions. Prior to computing the loss, both predicted and reference volumes were normalized to the range $[0, 1]$ to ensure numerical stability and comparability across samples.
The loss function is defined as follows:
\begin{align*}
\mathcal{L}_{total} = \mathcal{L}_{mse} + \gamma \cdot\mathcal{L}_{ssim},\tag{11}
\label{eq:myequation18}
\end{align*}

Parameter updates were performed using the AdamW optimizer with a one-cycle learning rate policy. The learning rate was cycled between 0.2 and 2 over 500 epochs, enabling faster convergence. Since only the redundancy weights were optimized, this comparatively large range was stable in practice.

\section{Results}
\label{sec:5}

\begin{figure*}[]
\centering
\includegraphics[width=7.5in]{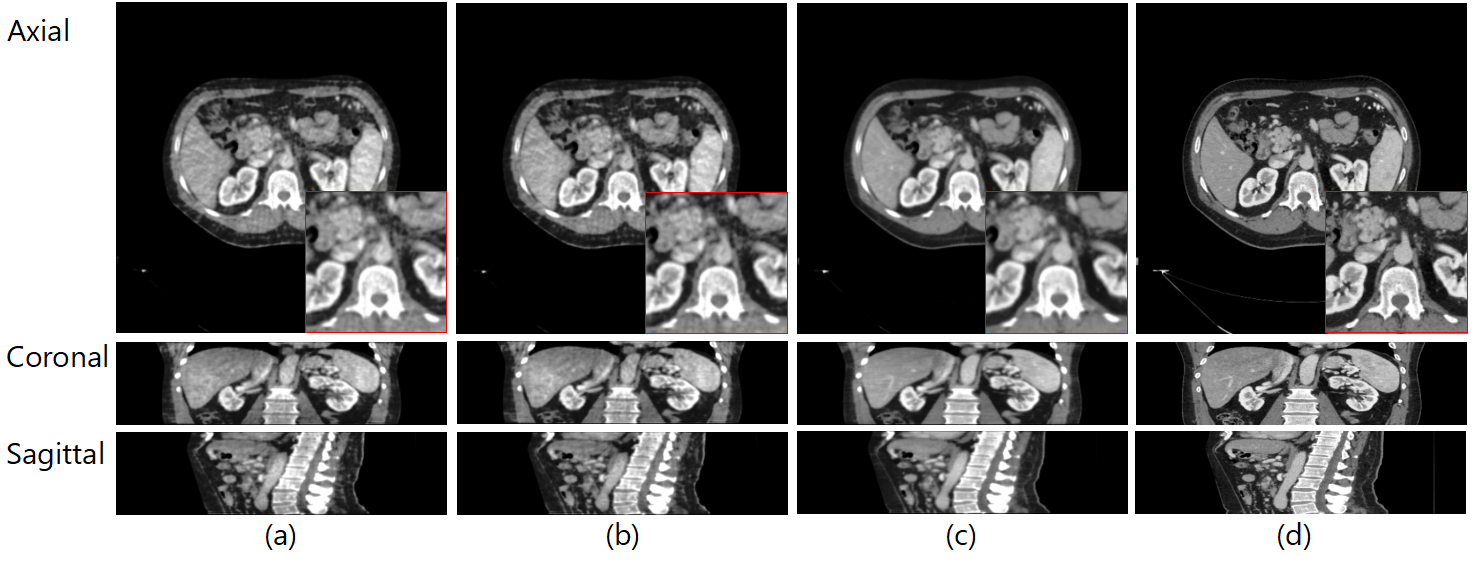}
\caption{Representative reconstructions for different trajectory types (Seed 0). Visualization parameters: window width = 400, window level = 60, both in Hounsfield Units (HU). (a) RNNR trajectory. (b) RFNR trajectory. (c) Sinusoidal trajectory. (d) Ground truth.}
\label{3684-result}
\end{figure*}

This section presents the experimental results along three dimensions: fixed-isocenter trajectory robustness (Section~\ref{sec:traj-results}), sparse-view reconstruction performance (Section~\ref{sec:sparse-results}), and evaluation on a multi-isocenter geometry (Section~\ref{sec:lissajous}).

\subsection{Fixed-Isocenter Trajectory Robustness}
\label{sec:traj-results}

In order to comprehensively assess the reconstruction performance under varying trajectory types, experiments were conducted on RT, RNNR, and RFNR trajectories, respectively. For each trajectory type, five distinct geometries were generated using different random seeds, and for each geometry, five test samples were created, resulting in 25 test cases per trajectory type and 75 in total, as described in Section~\ref{sec:data-preparation}. Quantitative evaluation metrics, namely MSE, peak signal-to-noise ratio (PSNR), and SSIM, were calculated for each case.

The quantitative results are summarized in Table~\ref{tab:Trajectory}. For each trajectory type, the reported values correspond to the mean and standard deviation over the five trajectory instances generated from different random seeds.

Across all three trajectory types, the results are highly consistent across random seeds, indicating stable reconstruction performance of differentiable SV-FBP under different fixed-isocenter sampling patterns. In particular, even under the more aggressive RFNR ordering, the framework achieves reconstruction quality comparable to that of RT and RNNR. This suggests that, within the tested trajectory classes, reconstruction quality is not primarily determined by trajectory ordering or continuity alone.

In Table~\ref{tab:Trajectory}, $N/A$ indicates that analytical SV-FBP is not applicable to the discrete trajectories considered here. Since the analytical formulation relies on continuous trajectory derivatives and more regular sampling assumptions, it cannot be directly applied to the irregular and non-uniform trajectories used in this study.

Figure~\ref{3684-result} provides representative visual examples for the different trajectory types and shows that high-quality reconstructions are obtained across RT, RNNR, and RFNR despite the differing degrees of trajectory discontinuity.

For comparison, our previous work~\citep{ye2025draco} reported an MSE of 0.0904 $\pm$ 0.0149, a PSNR of 37.20 $\pm$ 1.34, and an SSIM of 0.9591 $\pm$ 0.0051 for a continuous sinusoidal trajectory. Compared with those results, the present discontinuous fixed-isocenter trajectories lead to a moderate reduction in reconstruction quality, which is consistent with the increased sampling irregularity. This trend is also visible in Fig.~\ref{3684-result}, where the discontinuous trajectories exhibit more pronounced streaking artifacts than the continuous sinusoidal reference.

\begin{figure}[t]
\centering
\includegraphics[width=3in]{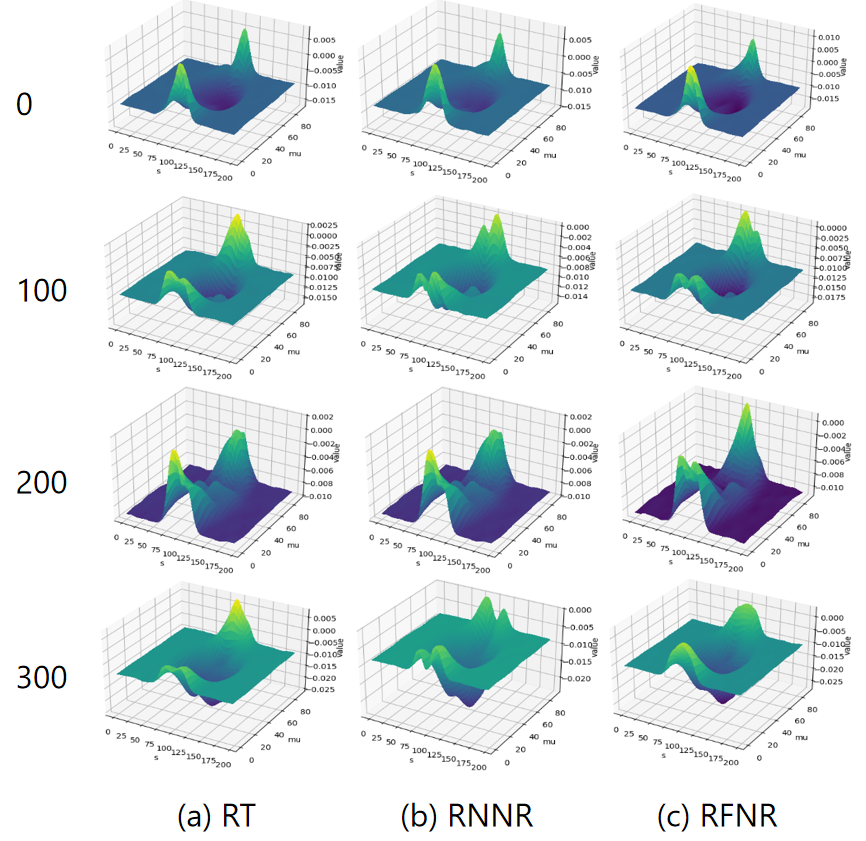}
\caption{Visualization of learned redundancy weights for different fixed-isocenter trajectory classes after restoring RNNR and RFNR to the same source-position indexing as RT. Each row corresponds to one representative projection index ($0$, $100$, $200$, and $300$), and each column corresponds to one trajectory type. After alignment, the learned weights exhibit highly similar spatial patterns across RT, RNNR, and RFNR, with differences mainly confined to local amplitude variations.}
\label{fig:weight_compare}
\end{figure}

To further investigate the behavior of the learned redundancy weights across different fixed-isocenter trajectory classes, we additionally visualized representative weights for RT, RNNR, and RFNR after restoring the reordered trajectories to the same source-position indexing as RT. Figure~\ref{fig:weight_compare} shows four representative projection indices. After this alignment, the learned weights exhibit highly similar spatial patterns across the three trajectory classes, with corresponding peaks, valleys, and smooth transitions appearing at comparable locations. The remaining differences are mainly local and do not change the overall structure of the learned weighting behavior.

\subsection{Sparse-View Reconstruction Analysis}
\label{sec:sparse-results}

\begin{table*}[]
\centering
\caption{Quantitative results for sparse-view reconstruction.}
%\resizebox{\textwidth}{!}{
\begin{tabular}{@{}lcccccc@{}}
\hline
\toprule

\emph{Algorithm} & \emph{\#Projections} & \emph{MSE} $\downarrow$ & \emph{PSNR (dB)} $\uparrow$ & \emph{SSIM} $\uparrow$ \\ 
%\hline
\midrule

\multirow{4}{*}{\emph{SV-FBP}} & 400 & 0.1041 ± 0.0160 & 35.95 ± 1.25 & 0.9208 ± 0.0079 \\
&300 & 0.1145 ± 0.0183 & 35.14 ± 1.26 & 0.9056 ± 0.0081\\
&200 & 0.1253 ± 0.0204 & 34.36 ± 1.30 & 0.8859 ± 0.0106\\
&100 & 0.1551 ± 0.0265 & 32.52 ± 1.29 & 0.8353 ± 0.0143\\
\hline
\multirow{4}{*}{\emph{AIR(50 iterations)}} & 400 & 0.1226 ± 0.0338 & 34.74 ± 1.52 & 0.9140 ± 0.0210 \\
&300 & 0.1235 ± 0.0240 & 34.53 ± 1.18 & 0.9175 ± 0.0115\\
&200 & 0.1315 ± 0.0255 & 33.99 ± 1.37 & 0.9119 ± 0.0121\\
&100 & 0.1507 ± 0.0285 & 32.79 ± 1.13 & 0.8875 ± 0.0084\\

\bottomrule

\end{tabular}%}%
\label{tab:Sparse}
\end{table*}

Because sparse-view reconstruction is of practical interest for reducing radiation dose and scan time, we further evaluated differentiable SV-FBP under progressively reduced projection counts. As a reference baseline, we included an algebraic iterative reconstruction (AIR) method with 50 iterations. AIR was implemented using the same PyroNN library and without additional regularization, ensuring a controlled comparison.

Table~\ref{tab:Sparse} summarizes the quantitative results for 400, 300, 200, and 100 projections. As expected, both methods show a gradual decline in reconstruction quality as the number of projections decreases. At moderate sampling densities (300--400 views), differentiable SV-FBP remains competitive with AIR and achieves comparable or slightly better PSNR and SSIM while reducing reconstruction time by approximately one order of magnitude. This indicates that the practical operating range of differentiable SV-FBP extends well into moderately undersampled settings.

Under more severe undersampling (100--200 views), however, the reconstruction quality of SV-FBP degrades more noticeably and gradually falls behind the iterative baseline. This transition is consistent with the absence of iterative data consistency in the SV-FBP pipeline, whereas AIR can partially compensate for sparse or inconsistent measurements through repeated updates. These results therefore identify a clear data-sufficiency boundary: in the present setup, differentiable SV-FBP remains competitive at 300--400 views but becomes less reliable under stronger undersampling.

Although AIR would likely improve further with more iterations, our previous study~\citep{ye2025draco} showed that differentiable SV-FBP can approach the performance of AIR with 100 iterations when trained with a lower learning rate and longer optimization. In the present study, AIR with 50 iterations was selected as a practical baseline to balance reconstruction quality and computational cost, since our focus is on the effect of trajectory complexity and sampling density rather than on exhaustive optimization of the iterative method.

\begin{figure*}[t]
\centering
\includegraphics[width=7in]{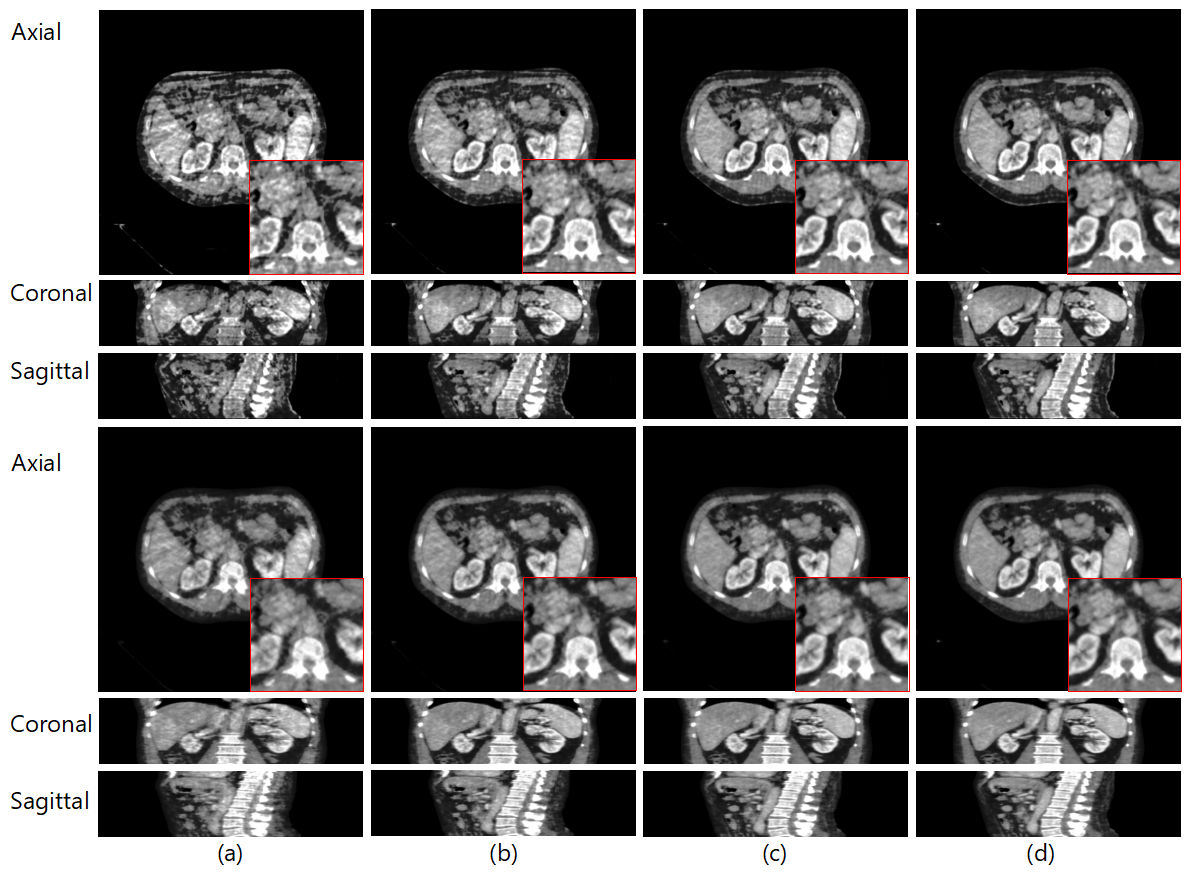}
\caption{
Representative sparse-view reconstructions using the RFNR trajectory (Seed 0). Visualization parameters: window width = 400 and window level = 60, both in Hounsfield Units (HU). The top row shows reconstructions obtained with differentiable SV-FBP, and the bottom row shows AIR (50 iterations). (a) 100 projections. (b) 200 projections. (c) 300 projections. (d) 400 projections.}
\label{fig:sparse_result}
\end{figure*}

To complement the quantitative results, Figure~\ref{fig:sparse_result} shows representative reconstructions for 100, 200, 300, and 400 projections using the RFNR trajectory (Seed 0). Across all projection counts, SV-FBP preserves comparatively sharp structural boundaries. As the number of projections decreases, both methods exhibit increasing artifacts and detail loss, consistent with the quantitative trends in Table~\ref{tab:Sparse}. Visually, AIR tends to suppress streaking artifacts more effectively under severe undersampling, but its reconstructions appear slightly smoother. In contrast, differentiable SV-FBP maintains clearer edges while showing more visible undersampling artifacts in the most challenging cases.

\subsection{Evaluation on a Multi-Isocenter Geometry}
\label{sec:lissajous}

The experiments above establish that the differentiable SV-FBP framework is robust to sampling irregularity and source ordering within fixed-isocenter geometries. To extend the evaluation beyond this setting, we further assess the framework on the Lissajous-saddle trajectory introduced in Section~\ref{sec:lissajous-def}, in which the rotation center varies continuously throughout the scan.

\begin{figure*}[!t]
\centering
\includegraphics[width=7.0in]{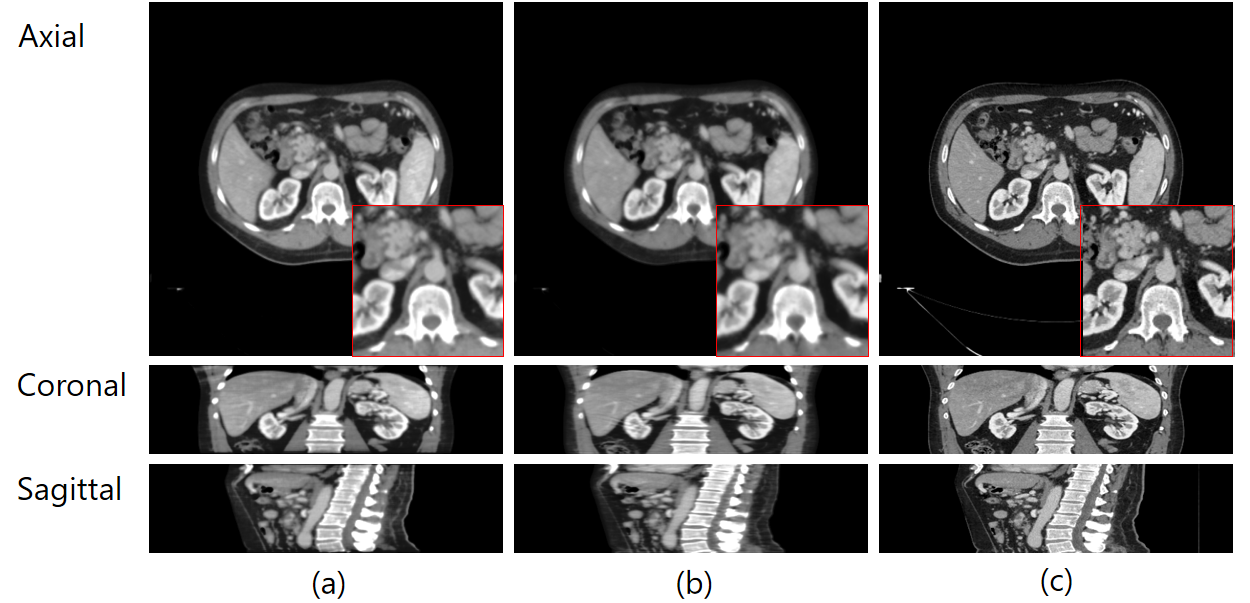}
\caption{Representative reconstructions for the Lissajous-saddle trajectory. (a) Differentiable SV-FBP. (b) AIR (50 iterations). (c) Ground truth.}
\label{fig:lissajous-result}
\end{figure*}

Quantitative results are summarized in Table~\ref{tab:lissajous}. The MSE, PSNR, and SSIM obtained for the Lissajous-saddle trajectory are comparable to those observed for the fixed-isocenter sinusoidal trajectory, indicating that the continuously varying isocenter does not substantially degrade reconstruction quality in the present setup. Representative reconstructions in Fig.~\ref{fig:lissajous-result} further show that fine structural details are recovered with only limited artifacts.

Taken together, these results show that differentiable SV-FBP remains applicable in a multi-isocenter acquisition setting without requiring architectural modification. This experiment therefore extends the robustness evaluation beyond the fixed-isocenter geometries considered in prior studies.

\begin{table}[htbp]
\centering
\small
\caption{Quantitative results for the Lissajous-saddle trajectory (mean $\pm$ standard deviation).}
\begin{tabular}{@{}lcccccc@{}}
\toprule
 & \textbf{MSE}$\downarrow$ & \textbf{PSNR (dB)}$\uparrow$ & \textbf{SSIM}$\uparrow$\\ 
\midrule
Our & 0.0984± 0.0127&  36.41± 1.33& 0.9521±0.0080 \\
AIR(50) & 0.1158± 0.0260&  35.16± 1.52& 0.9224±0.0292\\
\bottomrule
\end{tabular}%
\label{tab:lissajous}
\end{table}

\section{Discussion}
\label{sec:6}

\textbf{Trajectory continuity versus sampling distribution.}  
One of the main observations of this study is that trajectory continuity alone is not the dominant factor governing reconstruction quality. Although continuous trajectories such as the sinusoidal scan still provide the best overall performance, reordering randomly sampled source positions into a more continuous traversal (RNNR) does not lead to a substantial improvement over unreordered random trajectories. This suggests that, for a fixed set of sampling points, reconstruction quality is influenced more strongly by the spatial distribution of those points than by the continuity of the traversal path itself.

This interpretation is further supported by the analysis of the learned redundancy weights. After restoring RNNR and RFNR to the same source-position indexing as RT, the corresponding weight maps exhibit highly similar spatial patterns across the three trajectory classes. This indicates that different traversal orders mainly affect the ordering of the learned weights rather than their overall structure. In other words, continuity can be beneficial, but it cannot compensate for information loss caused by irregular or sparse spatial sampling. This distinction is important when interpreting the robustness of differentiable SV-FBP under discontinuous trajectories.

\textbf{Operating range under sparse-view conditions.}  
The sparse-view experiments provide a clearer picture of the data-sufficiency conditions under which differentiable SV-FBP remains practically useful. In the present setup, the method remains competitive with AIR at moderate sampling densities (300--400 projections) while offering substantially lower reconstruction time. Under stronger undersampling (100--200 projections), however, the performance gap in favor of the iterative baseline becomes more apparent. This behavior is consistent with the structure of the method: although differentiable SV-FBP benefits from learned redundancy weighting, it still follows an FBP-type pipeline and therefore lacks the iterative data-consistency mechanism that helps AIR remain stable in more ill-posed settings. The results therefore suggest a practical operating range in which differentiable SV-FBP combines efficiency with good reconstruction quality, while also making clear where its limitations begin to emerge.

\textbf{Evaluation beyond fixed-isocenter geometries.}  
The Lissajous-saddle experiment extends the evaluation beyond the fixed-isocenter setting used in the earlier robustness studies. In this geometry, the rotation center varies continuously throughout the scan and the source motion becomes non-planar and asymmetric. Despite this increased geometric complexity, the framework maintains reconstruction quality comparable to that obtained in the fixed-isocenter experiments. This indicates that the differentiable SV-FBP pipeline remains applicable in a multi-isocenter acquisition setting without requiring architectural modification. From a practical perspective, this is relevant for emerging robotic CBCT systems, where acquisition trajectories may depart substantially from standard circular or fixed-isocenter designs.

\textbf{Implications and future directions.}  
Taken together, these findings position differentiable SV-FBP as a useful compromise between the speed of analytical reconstruction and the flexibility of learned, geometry-aware weighting. At the same time, the experiments also clarify its current limitations: the method is robust to strong trajectory irregularity, but remains sensitive to severe undersampling. A natural direction for future work is therefore to combine the present framework with explicit data-consistency mechanisms, plug-and-play priors, or other hybrid strategies that preserve the efficiency and interpretability of differentiable SV-FBP while improving its performance in more ill-posed acquisition scenarios.

\section{Conclusions}
\label{sec:7}

This paper presents a systematic evaluation of a differentiable SV-FBP framework under challenging CBCT acquisition settings, including discontinuous fixed-isocenter trajectories, sparse-view sampling, and a multi-isocenter geometry. The experiments show that the framework maintains stable reconstruction performance across different fixed-isocenter trajectory orderings, indicating that reconstruction quality is influenced more strongly by the spatial distribution of sampling points than by trajectory continuity alone. Under sparse-view conditions, differentiable SV-FBP remains competitive with an iterative baseline at moderate sampling densities while providing substantially faster reconstruction, but its performance degrades more noticeably under severe undersampling.

In addition, the evaluation on a Lissajous-saddle trajectory shows that the framework remains applicable in a multi-isocenter acquisition setting without requiring architectural modification. Taken together, these results clarify both the robustness and the practical limits of differentiable SV-FBP. The framework offers a useful compromise between the efficiency of analytical reconstruction and the flexibility of data-driven redundancy weighting, making it a promising option for non-standard and robotic CBCT acquisition scenarios.

%%%%%%%%%%%%%%%%%%%%%%%%%%%%%%%%%%%%%%%%%%%%%%%%%%%%%%%%%%%%%%%%%%%%%%%
% Mandatory Sections. Please complete, especially for final publication
%%%%%%%%%%%%%%%%%%%%%%%%%%%%%%%%%%%%%%%%%%%%%%%%%%%%%%%%%%%%%%%%%%%%%%%

% Acknowledgements.
% Please include any funding, intellectual contributions not included in the authorship, and any other acknowledgements.
\acks{This work was supported by the Deutsche Forschungsgemeinschaft (DFG, German Research Foundation) through the project ``Deep-Learning-basierte CBCT-Rekonstruktion beliebiger Trajektorien'' (MA~4898/32-1). 

The authors gratefully acknowledge the scientific support and HPC resources provided by the Erlangen National High Performance Computing Center (NHR@FAU) of the FAU Erlangen-Nürnberg. }

% Ethical Standards.
% Please edit with the appropriate ethics considerations for your work. Include any pertinent IRB information, etc.
%
% Please note that the submission requirements included:
% The work presented must follow appropriate ethical standards in conducting research and writing the manuscript, following all applicable laws and regulations regarding treatment of animals or human subjects.
\ethics{The work follows appropriate ethical standards in conducting research and writing the manuscript, following all applicable laws and regulations regarding treatment of animals or human subjects.}

% Conflict of Interest
% Declaration of possible conflicts of interest: Authors must disclose any financial, organisational, commercial or personal conflicts of interest that might bias their work.
% If no conflicts, please say "We declare we don't have conflicts of interest."
\coi{We declare we don't have conflicts of interest.}

% Data availability
\data{The data supporting the findings of this study are available within the article and its supplementary materials.}

\bibliography{sample}

@article{hatamikia2022source,
  title={Source-detector trajectory optimization in cone-beam computed tomography: a comprehensive review on today’s state-of-the-art},
  author={Hatamikia, S and Biguri, A and Herl, G and Kronreif, G and Reynolds, T and Kettenbach, J and Russ, T and Tersol, A and Maier, A and Figl, M and others},
  journal={Physics in Medicine \& Biology},
  volume={67},
  number={16},
  pages={16TR03},
  year={2022},
  publisher={IOP Publishing}
}

@inproceedings{gang2020metal,
  title={Metal-tolerant noncircular orbit design and implementation on robotic c-arm systems},
  author={Gang, Grace J and Russ, Tom and Ma, Yiqun and Toennes, Christian and Siewerdsen, Jeffrey H and Schad, Lothar R and Stayman, J Webster},
  booktitle={Conference proceedings. International Conference on Image Formation in X-Ray Computed Tomography},
  volume={2020},
  pages={400},
  year={2020}
}

@inproceedings{gang2020non,
  title={Non-circular CT orbit design for elimination of metal artifacts},
  author={Gang, Grace J and Siewerdsen, Jeffrey H and Stayman, J Webster},
  booktitle={Medical imaging 2020: physics of medical imaging},
  volume={11312},
  pages={531--536},
  year={2020},
  organization={SPIE}
}

@article{herl2020scanning,
  title={Scanning trajectory optimisation using a quantitative Tuybased local quality estimation for robot-based X-ray computed tomography},
  author={Herl, Gabriel and Hiller, Jochen and Maier, Andreas},
  journal={Nondestructive Testing and Evaluation},
  volume={35},
  number={3},
  pages={287--303},
  year={2020},
  publisher={Taylor \& Francis}
}

@article{pack2004investigation,
  title={Investigation of saddle trajectories for cardiac CT imaging in cone-beam geometry},
  author={Pack, Jed D and Noo, Fr{\'e}d{\'e}ric and Kudo, H},
  journal={Physics in Medicine \& Biology},
  volume={49},
  number={11},
  pages={2317},
  year={2004},
  publisher={IOP Publishing}
}

@inproceedings{ye2025learned,
  title={Learned Shift-variant CBCT Reconstruction Weights for Non-continuous Trajectories},
  author={Ye, Chengze and Schneider, Linda-Sophie and Sun, Yipen and Thies, Mareike and Maier, Andreas},
  booktitle={BVM Workshop},
  pages={292--297},
  year={2025},
  organization={Springer}
}

@article{feldkamp1984practical,
  title={Practical cone-beam algorithm},
  author={Feldkamp, Lee A and Davis, Lloyd C and Kress, James W},
  journal={Journal of the Optical Society of America A},
  volume={1},
  number={6},
  pages={612--619},
  year={1984},
  publisher={Optical Society of America}
}

@article{liu2014model,
  title={Model-based iterative reconstruction: a promising algorithm for today's computed tomography imaging},
  author={Liu, Lu},
  journal={Journal of Medical imaging and Radiation sciences},
  volume={45},
  number={2},
  pages={131--136},
  year={2014},
  publisher={Elsevier}
}

@inproceedings{grangeat1991mathematical,
  title={Mathematical framework of cone beam 3D reconstruction via the first derivative of the Radon transform},
  author={Grangeat, Pierre},
  booktitle={Mathematical Methods in Tomography: Proceedings of a Conference held in Oberwolfach, Germany, 5--11 June, 1990},
  pages={66--97},
  year={1991},
  organization={Springer}
}

@article{li2023sparse,
  title={Sparse-view ct reconstruction with 3d gaussian volumetric representation},
  author={Li, Yingtai and Fu, Xueming and Zhao, Shang and Jin, Ruiyang and Zhou, S Kevin},
  journal={arXiv preprint arXiv:2312.15676},
  year={2023}
}

@inproceedings{lin2024learning,
  title={Learning 3D Gaussians for extremely sparse-view cone-beam CT reconstruction},
  author={Lin, Yiqun and Wang, Hualiang and Chen, Jixiang and Li, Xiaomeng},
  booktitle={International Conference on Medical Image Computing and Computer-Assisted Intervention},
  pages={425--435},
  year={2024},
  organization={Springer}
}

@article{wu2024differentiable,
  title={Differentiable Gaussian Representation for Incomplete CT Reconstruction},
  author={Wu, Shaokai and Lu, Yuxiang and Ji, Wei and Huang, Suizhi and Yang, Fengyu and Sirejiding, Shalayiding and He, Qichen and Tong, Jing and Ji, Yanbiao and Ding, Yue and others},
  journal={arXiv preprint arXiv:2411.04844},
  year={2024}
}

@article{wang2024neural,
  title={Neural Radiance Fields in Medical Imaging: Challenges and Next Steps},
  author={Wang, Xin and Hu, Shu and Fan, Heng and Zhu, Hongtu and Li, Xin},
  journal={arXiv preprint arXiv:2402.17797},
  year={2024}
}

@inproceedings{molaei2023implicit,
  title={Implicit neural representation in medical imaging: A comparative survey},
  author={Molaei, Amirali and Aminimehr, Amirhossein and Tavakoli, Armin and Kazerouni, Amirhossein and Azad, Bobby and Azad, Reza and Merhof, Dorit},
  booktitle={Proceedings of the IEEE/CVF International Conference on Computer Vision},
  pages={2381--2391},
  year={2023}
}

@article{defrise1994cone,
  title={A cone-beam reconstruction algorithm using shift-variant filtering and cone-beam backprojection},
  author={Defrise, Michel and Clack, Rolf},
  journal={IEEE transactions on medical imaging},
  volume={13},
  number={1},
  pages={186--195},
  year={1994},
  publisher={IEEE}
}

@article{ye2024deep,
  title={Deep Learning Computed Tomography based on the Defrise and Clack Algorithm},
  author={Ye, Chengze and Schneider, Linda-Sophie and Sun, Yipeng and Maier, Andreas},
  journal={arXiv preprint arXiv:2403.00426},
  year={2024}
}

@article{hong2021breath,
  title={Breath-hold CT and cone-beam CT images with expert manual organ-at-risk segmentations from radiation treatments of locally advanced pancreatic cancer [Data set]. The Cancer Imaging Archive},
  author={Hong, J and Reyngold, M and Crane, C and Cuaron, J and Hajj, C and Mann, J and Zinovoy, M and Yorke, E and LoCastro, E and Apte, AP and others},
  journal={The Cancer Imaging Archive https://doi. org/10.7937/TCIA. ESHQ-4D90},
  year={2021}
}

@article{syben2019pyro,
  title={PYRO-NN: Python reconstruction operators in neural networks},
  author={Syben, Christopher and Michen, Markus and Stimpel, Bernhard and Seitz, Stephan and Ploner, Stefan and Maier, Andreas K},
  journal={Medical physics},
  volume={46},
  number={11},
  pages={5110--5115},
  year={2019},
  publisher={Wiley Online Library}
}

@article{zhu2018image,
  title={Image reconstruction by domain-transform manifold learning},
  author={Zhu, Bo and Liu, Jeremiah Z and Cauley, Stephen F and Rosen, Bruce R and Rosen, Matthew S},
  journal={Nature},
  volume={555},
  number={7697},
  pages={487--492},
  year={2018},
  publisher={Nature Publishing Group UK London}
}

@article{he2020radon,
  title={Radon inversion via deep learning},
  author={He, Ji and Wang, Yongbo and Ma, Jianhua},
  journal={IEEE transactions on medical imaging},
  volume={39},
  number={6},
  pages={2076--2087},
  year={2020},
  publisher={IEEE}
}

@article{koetzier2023deep,
  title={Deep learning image reconstruction for CT: technical principles and clinical prospects},
  author={Koetzier, Lennart R and Mastrodicasa, Domenico and Szczykutowicz, Timothy P and van der Werf, Niels R and Wang, Adam S and Sandfort, Veit and van der Molen, Aart J and Fleischmann, Dominik and Willemink, Martin J},
  journal={Radiology},
  volume={306},
  number={3},
  pages={e221257},
  year={2023},
  publisher={Radiological Society of North America}
}

@article{maier2019learning,
  title={Learning with known operators reduces maximum error bounds},
  author={Maier, Andreas K and Syben, Christopher and Stimpel, Bernhard and W{\"u}rfl, Tobias and Hoffmann, Mathis and Schebesch, Frank and Fu, Weilin and Mill, Leonid and Kling, Lasse and Christiansen, Silke},
  journal={Nature machine intelligence},
  volume={1},
  number={8},
  pages={373--380},
  year={2019},
  publisher={Nature Publishing Group UK London}
}

@article{wurfl2018deep,
  title={Deep learning computed tomography: Learning projection-domain weights from image domain in limited angle problems},
  author={W{\"u}rfl, Tobias and Hoffmann, Mathis and Christlein, Vincent and Breininger, Katharina and Huang, Yixin and Unberath, Mathias and Maier, Andreas K},
  journal={IEEE transactions on medical imaging},
  volume={37},
  number={6},
  pages={1454--1463},
  year={2018},
  publisher={IEEE}
}

@misc{syben1807deriving,
  title={Deriving Neural Network Architectures using Precision Learning: Parallel-to-fan beam Conversion. GCPR 2018},
  author={Syben, Christopher and Stimpel, Bernhard and Lommen, Jonathan and W{\"u}rfl, Tobias and D{\"o}rfler, Arnd and Maier, Andreas},
  year={1807}
}

@article{sun2025learning,
  title={Learning Wavelet-Sparse FDK for 3D Cone-Beam CT Reconstruction},
  author={Sun, Yipeng and Schneider, Linda-Sophie and Ye, Chengze and Gu, Mingxuan and Mei, Siyuan and Bayer, Siming and Maier, Andreas},
  journal={arXiv preprint arXiv:2505.13579},
  year={2025}
}

@article{ye2025draco,
  title={DRACO: differentiable reconstruction for arbitrary CBCT orbits},
  author={Ye, Chengze and Schneider, Linda-Sophie and Sun, Yipeng and Thies, Mareike and Mei, Siyuan and Maier, Andreas},
  journal={Physics in Medicine \& Biology},
  volume={70},
  number={7},
  pages={075005},
  year={2025},
  publisher={IOP Publishing}
}

% needed in general before appendices.
% \newpage

% Appendix is optional
% \clearpage
% \appendix
% \section{Proof of the central theorem}
% 	In this appendix we prove the central theorem and present additional experimental results.
% 	\noindent

% 	{\noindent \em Remainder omitted in this sample. }

% \section{Appendix section}
% 	\subsection{Appendix subsection}
% 		\subsubsection{Appendix subsubsection}
% 			\paragraph{Appendix paragraph} Lorem ipsum dolor sit amet, consectetur adipisicing elit, sed do eiusmod
% 			tempor incididunt ut labore et dolore magna aliqua. Ut enim ad minim veniam,
% 			quis nostrud exercitation ullamco laboris nisi ut aliquip ex ea commodo
% 			consequat. Duis aute irure dolor in reprehenderit in voluptate velit esse
% 			cillum dolore eu fugiat nulla pariatur. Excepteur sint occaecat cupidatat non
% 			proident, sunt in culpa qui officia deserunt mollit anim id est laborum.

\end{document}